\documentclass[twocolumn,showpacs,amsmath,amssymb,superscriptaddress,aps,prl]{revtex4-2}

\usepackage{graphicx}
\usepackage{float}
\usepackage{blindtext}
\usepackage{epstopdf}
\usepackage{xcolor}
\usepackage{lineno}
\usepackage{amsmath}
\usepackage{todonotes}
\usepackage{braket}
\usepackage{chngcntr}
\usepackage{comment}
\usepackage{ifpdf}
\usepackage{graphicx}
\usepackage{dcolumn}
\usepackage{bm}

\renewcommand{\theequation}{\thesection.\arabic{equation}}
\newcounter{mnotecount}[section]
\renewcommand{\themnotecount}{\thesection.\arabic{mnotecount}}
\newcommand{\mnotex}[1]
{\protect{\stepcounter{mnotecount}}$^{\mbox{\footnotesize
$
\bullet$\themnotecount}}$ \marginpar{
\raggedright\tiny\em
$\!\!\!\!\!\!\,\bullet$\themnotecount: #1} }

\newcommand{\nk}[1]{\textcolor{black}{#1}}

\newcommand{\mo}[1]{\textcolor{black}{#1}}

\counterwithout{equation}{section}

\begin{document}

\title{Nonequilibrium control of thermal and mechanical changes in a levitated system}
\author{Markus Rademacher}
\altaffiliation{Present address: Department of Physics \& Astronomy, University College London, London WC1E 6BT, UK}
\email{m.rademacher.18@ucl.ac.uk}
\affiliation{Vienna Center for Quantum Science and Technology (VCQ), Faculty of Physics, University of Vienna, A-1090 Vienna, Austria}
\author{Michael Konopik}
\affiliation{Institute for Theoretical Physics I, University of Stuttgart, D-70550 Stuttgart, Germany}
\author{Maxime Debiossac}
\affiliation{Vienna Center for Quantum Science and Technology (VCQ), Faculty of Physics, University of Vienna, A-1090 Vienna, Austria}
\author{David Grass}
\altaffiliation{Present address: Department of Chemistry, Duke University, Durham, NC 27708, USA.}
\affiliation{Vienna Center for Quantum Science and Technology (VCQ), Faculty of Physics, University of Vienna, A-1090 Vienna, Austria}
\author{Eric Lutz}
\affiliation{Institute for Theoretical Physics I, University of Stuttgart, D-70550 Stuttgart, Germany}
\author{Nikolai Kiesel}
\email{nikolai.kiesel@univie.ac.at}
\affiliation{Vienna Center for Quantum Science and Technology (VCQ), Faculty of Physics, University of Vienna, A-1090 Vienna, Austria}


\begin{abstract}
Fluctuation theorems are fundamental extensions of the second law of thermodynamics for small nonequilibrium systems.
While work and heat are equally important forms of energy exchange, fluctuation relations have not been experimentally assessed for the generic situation of simultaneous mechanical and thermal changes. Thermal driving is indeed generally slow and more difficult to realize than mechanical driving. Here, we use feedback cooling techniques to implement fast and
controlled temperature variations of an underdamped levitated microparticle that are one order of magnitude faster than the equilibration time. Combining mechanical and thermal control, we verify the validity of a 
fluctuation theorem that accounts for both contributions, well beyond the range of linear response theory. Our results allow the investigation of general far-from-equilibrium processes in microscopic systems that involve fast mechanical and thermal changes at the same time.
\end{abstract}

\maketitle

Work and heat are two central quantities in  thermodynamics. The energy change related to mechanical driving, that is, the variation of a  system parameter such as the position of a piston, corresponds to work. On the other hand,  the energy change related to thermal driving, created by a temperature difference, is associated with heat \cite{pippard1964elements}. While both variables are deterministic in macroscopic systems, they become stochastic at the microscopic scale owing  to the presence of  thermal fluctuations \cite{jarzynski2011equalities,seifert2012stochastic}. In such systems, the second law  has been generalized in the form of fluctuation theorems that account for the effects of non negligible fluctuations \cite{jarzynski2011equalities,seifert2012stochastic}. Fluctuation relations reveal the universal laws that govern the properties of the random nonequilibrium entropy production. Their general validity arbitrarily far from thermal equilibrium makes them particularly useful in the study of nonequilibrium systems \cite{jarzynski2011equalities,seifert2012stochastic}.

Fluctuation theorems for mechanical driving \cite{jarzynski1997nonequilibrium,crooks1999entropy} have been extensively investigated experimentally in the past decades in numerous systems \cite{ciliberto2013fluctuations,ciliberto2017experiments}, ranging from biomolecules \cite{liphardt2002equilibrium,collin2005verification} and colloidal particles \cite{wang2002experimental,blickle2006thermodynamics} to mechanical \cite{douarche2006work} and electronic \cite{saira2012test} systems. By contrast, only relatively few experimental studies have been devoted to fluctuation relations for thermal driving \cite{jarzynski2004classical}, based,  for instance, on a varying bath temperature \cite{gomez2011heat} or a fixed temperature difference between two systems \cite{berut2016stationary,ciliberto2013heat}. This reflects the fact that thermal control is generally slow and more difficult to implement than mechanical control \cite{dubi2011colloquium}. The effective bath temperature has, for example,  been modulated using laser {absorption} \cite{gomez2011heat,blickle2012realization}, mechanical \cite{berut2016stationary} and electrical \cite{martinez2016brownian,rossnagel2016single,mar15a} random  forcing. However, no fluctuation theorem for {simultaneous} mechanical and thermal drivings has  been studied experimentally yet, despite its relevance in many areas where such  changes occur at the same time \cite{jarzynski1999microscopic,chelli2007generalization,chelli2007numerical,chatelain2007temperature,williams2008nonequilibrium,chelli2009nonequilibrium}, including  the   case of microscopic heat engines \cite{blickle2012realization,martinez2016brownian,rossnagel2016single}.  

\begin{figure}[t!]
    \centering
    \includegraphics[width=0.8\columnwidth]{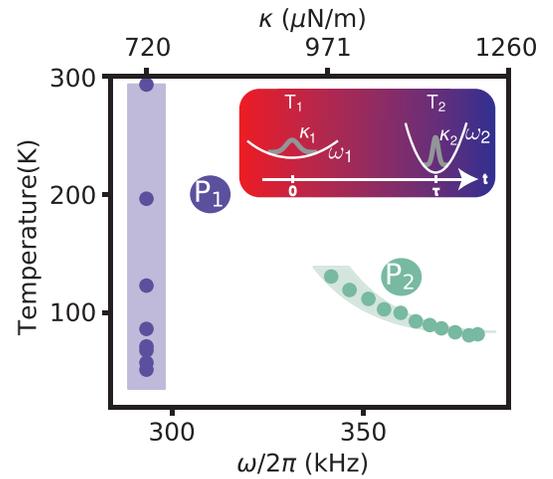}
    \caption{Nonequilibrium  processes with fast thermal and mechanical changes. We realize changes of the center-of-mass temperature $T$ and of the  spring constant $\kappa$ (corresponding to frequency $\omega$) of a harmonically trapped levitated microparticle by varying the feedback gain and the laser trap power (inset).  Protocol $P_1$ (purple) corresponds to a thermal change with fixed spring constant, while protocol $P_2$ (turquoise) refers to  simultaneous thermal and mechanical changes. Full circles are  data, with error bars smaller than the symbol size. Shaded areas represent the temperatures set experimentally, taking into account uncertainty due to laser power drifts. In the inset, the grey lines show position distributions during protocol  $P_2$ when the temperature is decreased from $T_1$ to $T_2$ and the frequency is simultaneously increased from $\omega_1$ to $\omega_2$ during a time $\tau$.}
    \label{fig:concept}
\end{figure}

\begin{figure*}[t]
    \centering
    \includegraphics[width=2\columnwidth]{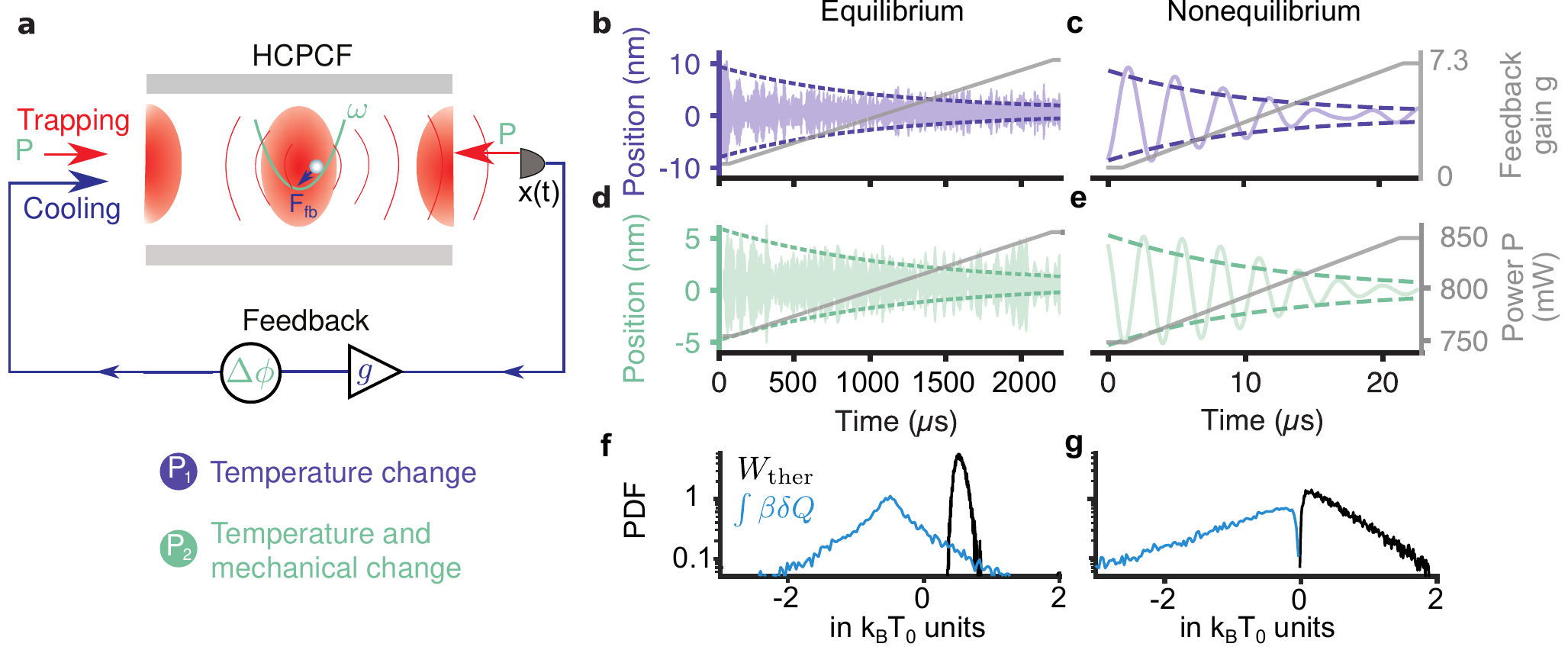}
    \caption{{Experimental setup and single trajectories. a)} Schematic of the experimental setup. Two counter-propagating  laser beams (red arrows) of wavelength $1064$ nm trap a 969 nm silica particle at the intensity maximum of the standing wave formed inside a hollow-core photonic crystal fiber (HCPCF).  The scattered light (red lines) along the HCFCP provides the center-of-mass motion $x(t)$ of the particle. An additional feedback laser beam (blue arrow) cools the center-of-mass motion of the particle. Here the velocity contribution of the feedback force $F_{\text{fb}}=g\gamma_\text{p} m \sin(\Delta \phi)\dot{x}$ depends on the feedback gain $g$ and the trapping laser power $P$ via the feedback phase $\Delta\phi$ (see Supplemental Material \cite{sup}).  Protocol $P_1$ (thermal change, purple) is implemented by a linear increase of the  feedback gain $g$ at constant optical trap power. Protocol $P_2$ (thermal and mechanical change, turquoise) is realized by changing the optical trap power $P$. {b-e)} Measured single-particle trajectories (solid lines) and ensemble variances (dashed lines, based on $15.000$ trajectories) of the particle center-of-mass motion are plotted together with the corresponding change of the respective control parameter (grey lines). Panels {b,d} ({c,e}) correspond to the slowest  equilibrium (fastest nonequilibrium) protocols, one order of magnitude slower (faster) than the relaxation time of the system. {f-g)} Experimental distributions of the dimensionless thermal work $W_\text{ther}$ (black) and dimensionless heat $\int\beta\delta Q$ (blue) in the case of protocol $P_2$, for slow (f) as well as fast (g) drivings.}
    
    \label{fig:traces}
\end{figure*}

\mo{In this paper}, we experimentally demonstrate \mo{fast} thermal and mechanical control of an oscillator \mo{on} timescales much \mo{shorter} than its relaxation time. We use our system to investigate \mo{for the first time} generalized fluctuation relations that account for \mo{simultaneous} mechanical and thermal changes \mo{far from equilibrium \cite{jarzynski1999microscopic,chelli2007generalization,chelli2007numerical,chatelain2007temperature,williams2008nonequilibrium,chelli2009nonequilibrium}}. 
\nk{To achieve this, we use optically trapped microparticles \cite{gieseler2018levitated,gieseler2014dynamic,hoang2018experimental}. Here, the harmonic trap created by the optical potential can be dynamically controlled and the effective environmental temperature of the particles center-of-mass motion can be set by linear feedback cooling \cite{li2010measurement}  (Fig.~\ref{fig:concept}).} \mo{While recent studies have examined the effect of information gain on the thermodynamics of a particle interacting with a constant temperature bath in the presence of feedback \cite{par15,toy10,kos14,rol14,rib19,debiossac2020thermodynamics}, we consider in the following  the dynamics of a particle coupled to a bath with an effective time-dependent temperature. Both situations are physically distinct and the corresponding fluctuation relations are fundamentally different.}
Specifically, we test two protocols (Fig.~\ref{fig:concept}): {first},  solely the  temperature  is modified (protocol $P_1$) and, {second}, both temperature and spring constant are varied (protocol $P_2$). With our approach, we are able to study nonequilibrium processes that occur one order of magnitude faster than the relaxation time of the system ($\gamma_\text{p}^{-1}$). In this regime, linear response theory is no longer applicable. Our findings emphasize the importance of the thermal (aka entropic) work, associated with a change of entropy of the bath \cite{crooks1999excursions}, on the same footing as the conventional mechanical work~\cite{sup}. Fluctuation theorems for thermal driving indeed only permit accurate determination of the equilibrium free energy difference when this novel contribution is included.

The generalized fluctuation relation for mechanical and thermal drivings of a system described by the Hamiltonian $H(\kappa)$, with time-dependent spring constant $\kappa(t)$, and inverse temperature $\beta(t)$  reads \cite{jarzynski1999microscopic,chelli2007generalization,chelli2007numerical,chatelain2007temperature,williams2008nonequilibrium,chelli2009nonequilibrium},
\begin{equation}
\left\langle \exp(-W)\right\rangle = \exp[-\Delta(\beta {F})].
\label{eq:WSE}
\end{equation} 
where $\Delta(\beta {F})=\beta(\tau) {F}(\tau)-\beta(0) {F}(0)$ with $ {F}$   the (equilibrium) free energy of the system. The  generalized dimensionless work $W$ is defined as \cite{chelli2007generalization,chelli2007numerical,chatelain2007temperature,williams2008nonequilibrium,chelli2009nonequilibrium},
\begin{equation}
W=W_\text{mech}+W_\text{ther}=\int_{0}^{\tau}dt\, \beta {\partial H}/{\partial \kappa}\,\dot{\kappa}+\int_{0}^{\tau}dt\,\dot{\beta}H.
\label{eq:W_general}
\end{equation} 
The first term in Eq.~\eqref{eq:W_general} corresponds to the dimensionless stochastic mechanical work along a single trajectory  of duration $\tau$, while the second term is  the random thermal (or entropic) work induced by a temperature change \cite{sup}. The brackets $\langle.\rangle$ denote an  average over many trajectories. The dimensionless heat exchanged with the bath follows from a generalized  first law  as $\int\beta\delta Q = \Delta(\beta H) - W$  with $\Delta(\beta {H})=\beta(\tau) {H}(\tau)-\beta(0) {H}(0)$ \cite{jarzynski1999microscopic}. Expression (\ref{eq:WSE}) reduces to the usual Jarzynski equality in the case of constant temperature, $\dot \beta =0$.

For a harmonic Hamiltonian, {$H(x,p,\kappa)={p^2}/{2m}+\kappa x^2/2$}, with position $x$, momentum $p$,  mass $m$ and spring constant $\kappa$, like in our experiment, the normalized free energy difference may be evaluated explicitly. We find
\begin{equation}
\Delta(\beta  {F})_\text{ho}=\ln[\beta(\tau)/\beta(0)]+\ln[\kappa(\tau)/\kappa(0)]/2.
\label{eqndbF}
\end{equation}
 The fluctuation theorem (1) holds arbitrarily far from equilibrium. Close to equilibrium, in the  linear response  regime, Eq.~(1) may be Taylor expanded \cite{jarzynski2011equalities,seifert2012stochastic}. To first order, one recovers the equilibrium result, $\Delta(\beta {F})_{\text{eq}}=\langle  {W}\rangle$, \mo{whereas}, to second order, one obtains the linear-response formula,  $\Delta(\beta {F})_{\text{lr}}= \langle  {W}\rangle-(\langle  {W}^2\rangle-\langle  {W}\rangle^2)/2$, where $\langle  {W}^2\rangle-\langle  {W}\rangle^2$ is the variance of the total work. 
 
 {The harmonic oscillator in our experiment is a levitated silica microsphere with a diameter of 969~nm. It is optically trapped by two counterpropagating laser beams at 1064~nm in the intensity maximum of a standing wave (Fig.~\ref{fig:traces}a) inside a hollow-core photonic crystal fiber (HCPCF)~\cite{grass2016optical}. The amplitude of the microsphere's center-of-mass motion is much smaller than the wavelength of the trapping laser, thus the optical potential is approximately harmonic with a tunable power-dependent frequency $\omega_\text{p}/2\pi$ between $250$~kHz and $400$~kHz. 
The gas surrounding the microsphere acts as a heat bath at room temperature. Its coupling to the microparticle's center-of-mass motion is determined by the pressure, which we set to 1.5~mbar. This results in a coupling rate of $\gamma_\text{p}/2\pi=6.3$~kHz, which is much smaller than the mechanical frequency. Hence, the underdamped Langevin equation for an harmonic oscillator applies \cite{sup}.} 

To perform work on the system and thus implement mechanical driving, we control the spring constant $\kappa$ via the optical trap power $P$, exploiting their linear dependence ($\kappa \propto P $). We further realize thermal control of the microparticle center-of-mass temperature via feedback control, i.e., cold-damping \cite{li2010measurement}. To implement feedback cooling, we exert a radiation pressure force from an additional laser that counteracts the particle motion. Specifically, we apply a delayed, linearly position-dependent force  $F_\text{fb}=-g m \gamma_\text{p}\omega_0 x(t-t_\text{fb})$ where $g$ is the feedback gain, $\omega_0= \sqrt{\kappa/m}$ the mechanical frequency {without feedback}, $m$ the mass of the microsphere, and $t_\text{fb}={5\pi}/{2\omega_0}$ the feedback delay {\cite{sup}}.  The feedback force may be split into a position and a velocity component \cite{sup}. The velocity component provides additional friction and therefore cools the temperature of the particle motion. The effective inverse temperature of the system is then given by {$\beta(t)=\beta_0 [ 1+g\sin\left(\Delta\phi\right)]$} where $\Delta\phi= \omega t_\text{fb}$ is the phase introduced by the delayed feedback, {with $\omega$ the mechanical frequency with feedback \cite{sup}}. We carry out protocol  $P_1$ (thermal driving) by increasing the gain $g(t)$ linearly in time to solely vary the temperature, while the {optical potential} remains unchanged. We furthermore  implement protocol $P_2$ (mechanical and thermal drivings) by  tuning the laser power of the trap, hence  the  mechanical frequency $\omega(t)$, and keeping the gain $g$  constant. This also changes the effective inverse bath temperature $\beta(t)$ due to the frequency dependence of the phase $\Delta\phi$.  In both cases, we measure the center-of-mass motion along the axis of the hollow-core  fiber  by interferometric readout of the light scattered by the microparticle \cite{sup,grass2016optical}. 

\begin{figure}[t!]
    \centering
    \includegraphics[width=0.9\columnwidth]{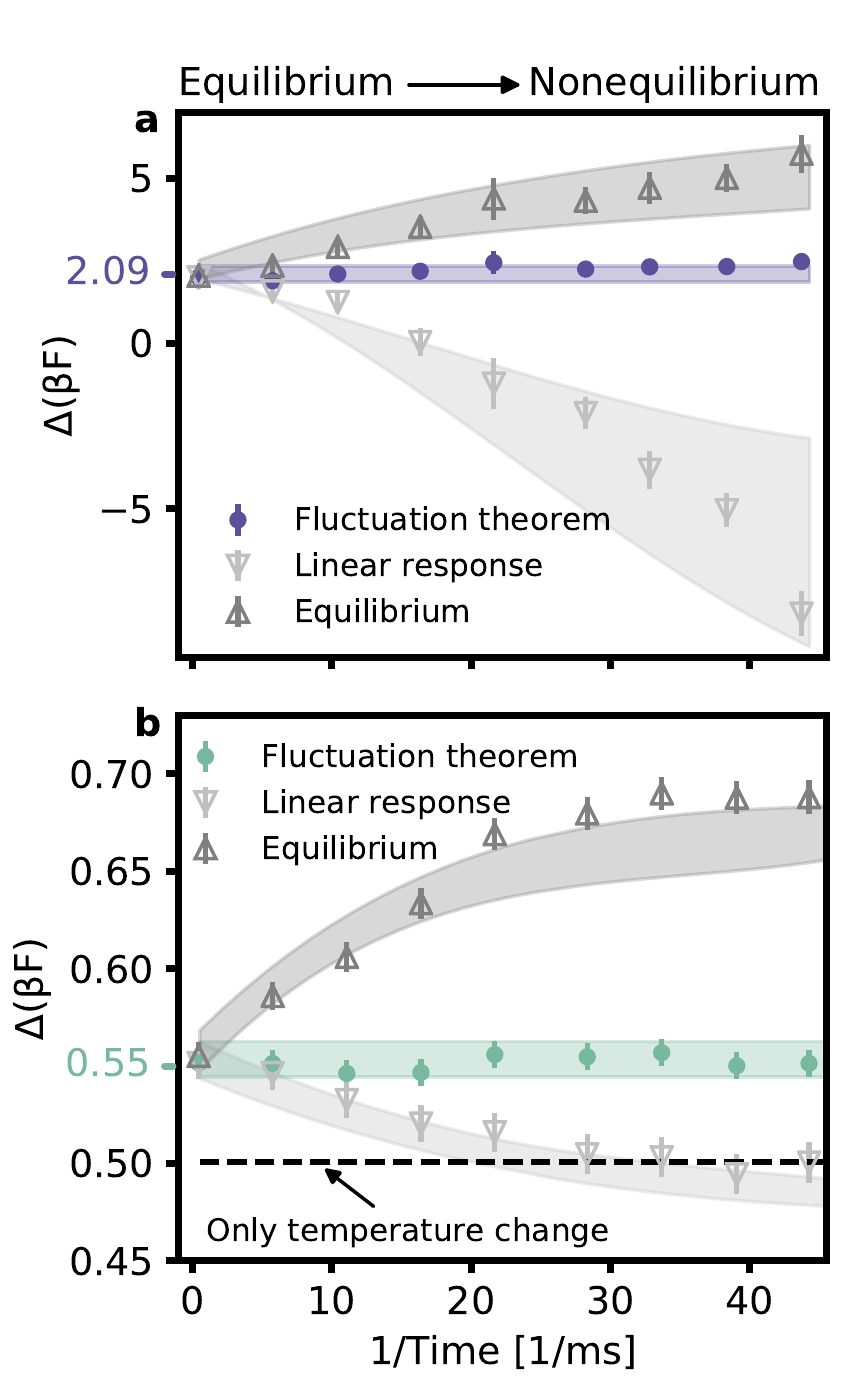}
    \caption{{Fluctuation theorem for thermal and mechanical modulations.} Normalized free energy difference $\Delta(\beta  {F})$ versus inverse driving  time $\tau$ for (a) thermal protocol $P_1$ and (b) thermal and mechanical protocol $P_2$. Dots represent experimental data evaluated using Eq.~\eqref{eq:WSE}.  Shaded areas are theoretical predictions with an uncertainty that incorporates long term laser drifts. Errors bars are determined by the standard deviation over 15,000  runs. 
   Equilibrium, $\Delta(\beta {F})_{\text{eq}}$, (triangle up) and linear response, $\Delta(\beta {F})_{\text{lr}}$, (triangle down) results only hold for slow driving. The horizontal dashed line in b) indicates the contribution of thermal change only.}
    \label{fig:WSE}
\end{figure}

Figures~\ref{fig:traces}(b-e) display recorded single-particle trajectories for protocols $P_1$ (purple) and $P_2$ (turquoise) for slow (equilibrium) and fast (nonequilibrium) drivings:  the solid (dashed) lines indicate the particle center-of-mass motion (ensemble variance). During  protocol $P_1$, the feedback gain $g$ is varied linearly from 0 to 7.3 (grey lines), decreasing the particle temperature by a factor of up to $\approx$ 8.3 (Fig.~\ref{fig:concept}). During protocol $P_2$, the laser trapping power $P$ is increased linearly from 750 to 850 mW, reducing the spring constant by $\approx 11$~\%, and decreasing the temperature by a factor $\approx$ 1.5 (Fig.~\ref{fig:concept}). Note that light absorption does not significantly influence the particle temperature in our case~\cite{sup}. Between each cycle and before the protocol ramp starts, the system equilibrates to the inverse effective temperature $\beta(0)$,  determined by the feedback.
  The duration $\tau$ of  equilibrium (nonequilibrium) protocols is 2.26 ms (22.6 $\mu$s). With $\gamma_\text{p}^{-1}\approx 0.2$ ms, equilibrium protocols thus probe the  particle dynamics in the quasistatic regime where $\tau\gg \gamma_\text{p}^{-1}$.  By contrast, nonequilibrium protocols are an order of magnitude faster than the relaxation time $ \tau  \ll \gamma_\text{p}^{-1}$ of the system.

\begin{figure}[t!]
    \centering
    \includegraphics[width=0.9\columnwidth]{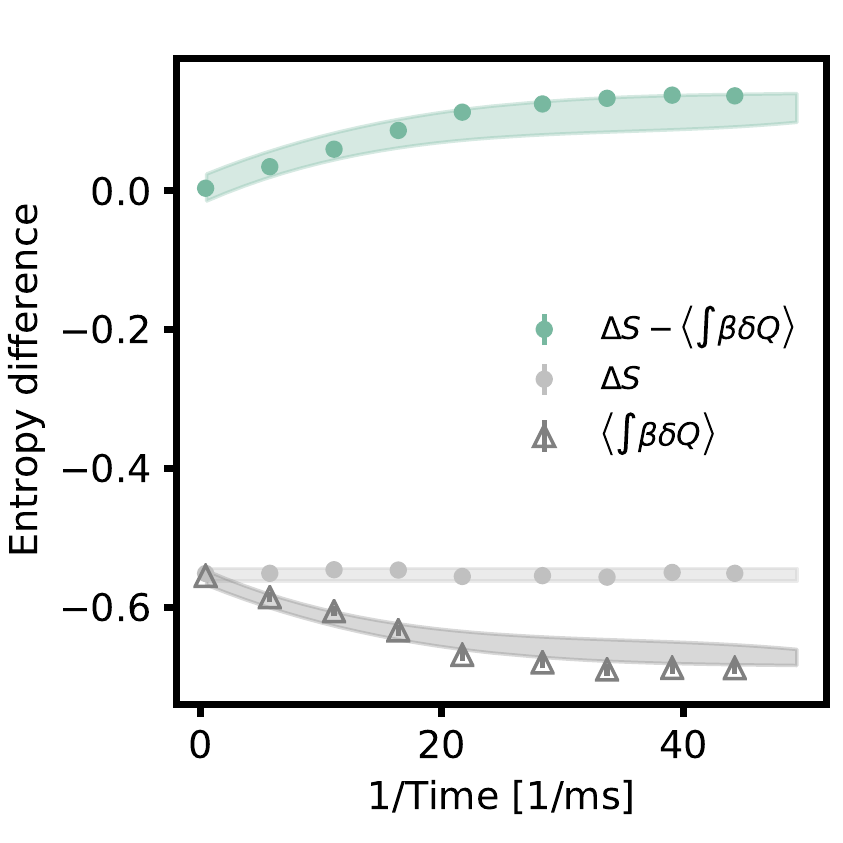}
    \caption{{Clausius inequality for thermal and mechanical changes.} Comparison between the measured (dimensionless) heat exchanged with the environment $\langle \int \beta \delta{Q}\rangle$ (triangles) and the (dimensionless) entropy variation  of the system $\Delta S$ (grey full circles), as a function of the inverse protocol time for  thermal and mechanical changes (protocol $P_2$) followed by an isothermal thermalization. The Clausius inequality, $ \Delta S -\langle \int \beta \delta Q\rangle \ge 0$,  is verified for any protocol speed. Shaded areas are theoretical predictions including the uncertainty on the underlying experimental parameters. Error bars are determined via the standard deviation of the respective value over 15,000  runs and are smaller than the symbol size.}
    \label{fig:Clausius}
\end{figure}
 
We evaluate  the random  mechanical  and thermal works,  ${W}_\text{mech}$ and $ {W}_\text{ther}$, together with the stochastic heat, $\int \beta \delta Q$, along 15,000 single trajectories. Figures~\ref{fig:traces}(f-g) show the distributions  of the dimensionless thermal work $W_\text{ther}$ (black) and dimensionless heat $\int \beta \delta Q$ (blue) for protocol $P_2$ (the corresponding distribution for ${W}_\text{mech}$ is presented in the Supplemental Material~\cite{sup}). For slow driving (f), the dimensionless thermal work has a Gaussian profile whereas the dimensionless heat distribution exhibits  exponential tails, in analogy to the case of pure mechanical driving \cite{joubaud2007fluctuation}. On the other hand, for fast  driving (g), the two distributions are clearly asymmetric. However, their exact shape is  yet unknown.

We further investigate, for both protocols, the normalized free energy difference $\Delta(\beta  {F})$ as a function of the inverse driving time associated with the speed of the protocol. We compare three methods to determine $\Delta(\beta  {F})$ from our experimental data: by applying the fluctuation theorem $\Delta(\beta {F})_{\text{ft}}$, Eq.~(1), (dots) by using the equilibrium result $\Delta(\beta {F})_{\text{eq}}$ (triangle up) and by employing the linear response formula $\Delta(\beta {F})_{\text{lr}}$ (triangle down). In addition, we show the calculated free energy difference $\Delta(\beta  {F})_\text{ho}$ as given by Eq.~(2).

For the purely thermal {control} implemented in protocol $P_1$ (Fig.~\ref{fig:WSE}a), we theoretically expect $\Delta(\beta {F})_\text{ho} = 2.09 \pm 0.24$. This value is inferred from equilibrium measurements of the system temperature and spring constant using Eq.~(2). \nk{Uncertainties on the theoretical prediction originate from slow drifts of these parameters and are shown as purple areas in the plot (for more detail, see \cite{sup})}.  We observe that the equilibrium  and linear response results,  $\Delta(\beta {F})_{\text{eq}}$  and $\Delta(\beta {F})_{\text{lr}}$, hold for long protocol times.  However, both  rapidly  deviate from $\Delta(\beta  {F})_\text{ho}$ for faster protocols. By contrast, we have very good agreement with the values obtained from the fluctuation relation  (1) (purple dots {in Fig.~\ref{fig:WSE}a}), within the error bars, for all driving speeds.  Even for the fastest protocol, we determine a value consistent with our expectation {$\Delta(\beta {F})_\text{ft} = 2.47 \pm 0.25$}.
Figure~\ref{fig:WSE}b shows similar results for the combined mechanical and thermal control implemented in protocol $P_2$: while the equilibrium and near-equilibrium approximations are valid for slow changes, the nonequilibrium fluctuation relation (1) (turquoise dots) correctly reproduces the equilibrium free energy difference $\Delta(\beta {F})_\text{ho} = 0.55 \pm 0.01$ (turquoise area), even very far from the quasistatic regime with {$\Delta(\beta {F})_\text{ft} = 0.55 \pm 0.01$} for the fastest protocol. The dashed black line represents the theoretical normalized free energy difference when only thermal work is taken into account. As it significantly departs from the actual value in the experiment, we conclude that our protocol actually requires accounting for both mechanical work and entropic work.

The  fluctuation relation (1) extends the Clausius statement of the second law  to stochastic far-from-equilibrium processes. Noting that $\Delta(\beta F) = \Delta (\beta H)_\text{eq} -\Delta S$, where $\Delta S$ is the (dimensionless) entropy change between initial  and final equilibrium states, and using the convexity of the exponential, we indeed have $\langle \int \beta  \delta Q \rangle \leq \Delta S$ \cite{jarzynski1999microscopic}. The integrated heat absorbed by the system, divided by the temperature at which that heat is absorbed, is thus bounded from above by the  entropy variation of the system. Figure  \ref{fig:Clausius} presents  the first experimental confirmation of that fundamental inequality in microscopic systems for protocol $P_2$ followed by an isothermal equilibration, for varying protocol speed. The inequality  is verified for arbitrarily far-from-equilibrium processes. Contrary to the generalized fluctuation theorem (1), which holds exactly as an equality, the Clausius inequality provides a worse bound for increasing protocol speeds.

In conclusion, we have  demonstrated an experimental route to implement fast and controlled temperature changes in an underdamped levitated harmonic system using feedback cooling techniques. We have exploited the ability to realize simultaneous mechanical and thermal drivings on timescales much shorter than the equilibration time of the system, to reveal the importance of the entropic work and extend the  applicability of  fluctuation theorems, for both mechanical and entropic works, beyond the  linear response regime. Our versatile experimental approach enables  the  study of generic nonequilibrium transformations in microscopic systems that involve fast mechanical and thermal modulations at the same time. Given the experimental access to the quantum regime recently achieved with levitated nanoparticles \cite{delic2020cooling, magrini20, tebbenjohanns21}, the extension of such fluctuation relations to the quantum domain appears to be an exciting prospect.

\begin{acknowledgements}
\section{Acknowledgements}
We thank Markus Aspelmeyer for his support and valuable discussions. N. K. acknowledges support from the Austrian Science Fund (FWF): Y 952-N36, START. D.G. acknowledges support through the doctoral school CoQuS (Project W1210). We further acknowledge financial support from the German Science Foundation (DFG) under project FOR 2724. We also acknowledge the support of NVIDIA Corporation with the donation of the Titan Xp GPU used for this research.
\end{acknowledgements}

\bibliographystyle{apsrev4-2}


%

\clearpage
\newpage
\widetext

\begin{center}
\textbf{\large Supplemental Material: Nonequilibrium control of thermal and mechanical changes in a levitated system}
\end{center}
\setcounter{equation}{0}
\setcounter{figure}{0}
\setcounter{table}{0}
\setcounter{page}{1}
\makeatletter
\renewcommand{\figurename}{Supplementary Figure}
\renewcommand{\theequation}{S\arabic{equation}}
\renewcommand{\thefigure}{S\arabic{figure}}
\renewcommand{\bibnumfmt}[1]{[S#1]}
\renewcommand{\citenumfont}[1]{S#1}
{Rademacher et al.}

\setcounter{equation}{0}
\setcounter{figure}{0}
\setcounter{table}{0}
\setcounter{page}{1}
\makeatletter
\renewcommand{\figurename}{Supplementary Figure}
\renewcommand{\theequation}{S\arabic{equation}}
\renewcommand{\thefigure}{S\arabic{figure}}
\renewcommand{\bibnumfmt}[1]{[S#1]}
\renewcommand{\citenumfont}[1]{S#1}

\onecolumngrid
\section{Langevin equation} 
 The dynamics of the center-of-mass motion of the harmonically bound levitated particle follows the underdamped Langevin equation  
\begin{align}
\label{eq:thermal-mech-lang}
\ddot{x} + \gamma_\text{p}\dot{x} + \omega_0^2 x- g\gamma_\text{p} \omega_0 x({t-t_\text{fb}}) = F_\text{th}/m,
\end{align}
where $g=\gamma_\text{fb}/\gamma_\text{p}$ is the feedback gain and $F_\text{th}$ denotes  a centered, delta-correlated Gaussian thermal noise force.  In the high-quality-factor approximation ($\omega_0 \gg \gamma_\text{p}$), the delayed position of the particle can be written as ( see also \cite{debiossac2020thermodynamics}) $x(t-t_\text{fb})=x\cos(\Delta\phi)-\sin(\Delta\phi)\dot{x}/\omega$ with $\Delta\phi=t_\text{fb}\omega$ the phase introduced by the delayed feedback and $t_\text{fb}=\frac{5\pi}{2\omega_0}$. The velocity component leads to an extra damping, $\gamma_\text{fb}=g\gamma_\text{p}\sin\left(\Delta\phi\right)$, while the position component to an optical spring effect with ${\omega_\text{fb}^2}=g\gamma_\text{p}\omega_0 \cos(\Delta\phi)$.  The effective temperature of the center-of-mass motion of the microparticle is defined as $T/T_0=(\gamma_\text{p}+\gamma_\text{fb})/\gamma_\text{p}$, which gives $\beta/\beta(0)=1+g\sin(\Delta\phi)$ with $\beta = 1/(k_B T)$ and $k_B$ the Boltzmann constant.
\bigskip

\section{Experimental setup} 

A 969 nm diameter silica particle is trapped at the intensity maximum of a standing wave inside a HCPCF, see Ref.~\cite{grass2016optical} and Supplementary Information. The trapping laser power is $(860 \pm 11)$ mW and the experiments were carried out at pressures of $1.5$ mbar and $3.5$ mbar. The particle center-of-mass motion is recorded using a balanced photodiode (Thorlabs, PDB425C-AC) by collecting $\sim$ 10\% of the light transmitted through the HCPCF and the light scattered by the particle. In protocol $P_1$, thermal change is achieved via velocity-correlated radiation pressure from a second laser beam which is orthogonally polarized and frequency shifted with respect to the trapping laser. The read-out signal of the center-of-mass motion is first bandpass filtered (f = $\omega_0/2\pi$, BW = 600 kHz), then delayed by $t_{\text{fb}}={5\pi}/{2\omega_0}$ with a field-programmable gate array (National Instrument, PXIe-7965). This time delay corresponds to the feedback cooling of the center-of-mass motion of the particle~\cite{debiossac2020thermodynamics}. Here $\omega_0$ is the mechanical frequency with the feedback beam turned off. The delayed signal is then amplified with a variable gain $g(t)$ and fed to the feedback AOM (IntraAction DE-805M) for power modulation. This results in a feedback force applied to the particle motion through radiation pressure. For protocol $P_2$, the feedback beam is used with a fixed gain and a delay $t_{\text{fb}}={5\pi}/{2\omega_0}$. In this case, the trapping laser power is modulated using an external ramp signal fed to both trapping AOM drivers (Moglabs XRF421).

\section{Stochastic work}

For each protocol, the work is computed from the measured positions   of the microparticle.\\

\textit{Protocol $P_1$}:
We evaluate the thermal work for protocol $P_1$ in Eq.~(1) for each protocol ramp using  the discrete form, $ {W}_\text{ther}=\sum_{i=1}^{N}(\beta_{i+1}-\beta_i) H_i$ with $N$ the number of points in each ramp. Here $\beta_i=\beta_0+(\beta_1-\beta_0)i/\tau$ is the thermal driving ramp. The discrete Hamiltonian is given by $H_i=(m\dot{x}_i^2+\kappa x_i^2)/2$. \\

\textit{Protocol $P_2$}: The total work for protocol $P_2$  is similarly computed as $ {W}=\sum_{i=1}^{N}[\beta_i(H_{i+1}-H_i) + (\beta_{i+1}-\beta_i) H_i]$. We have $\beta_i=1-2.55\sin(5\pi/2a_i)$ with the linear ramp $a_i=a_0+(a_1-a_0)i/\tau$ and $a_0=\omega(0)/\omega_0=0.9$ and $a_1=\omega(\tau)/\omega_0=1$. The  discrete Hamiltonian is in this case $H_i=(m\dot{x}_i^2+\kappa_i x_i^2)/2$ with $\kappa_i=\kappa a_i$. Note that the second term of the work may  be rewritten as $\beta_i(H_{i+1}-H_i)=\beta_i\kappa(a_1-a_0)x_i^2/\tau$, {since we apply mechanical work only by modifying the potential energy component of the total particle energy}.

\section{Temperature settings during both protocols} 

We here confirm the consistency between the expected and the experimentally determined effective bath temperature for our protocols ($P_1$ and $P_2$). We do so for the quasistatic cases ($\tau=2.26$~ms) of both protocols, as faster instances of the protocols do not allow such a direct comparison, because the particle motion has no time to equilibrate. \\

\begin{figure*}
\centering
\includegraphics[width=0.9\textwidth]{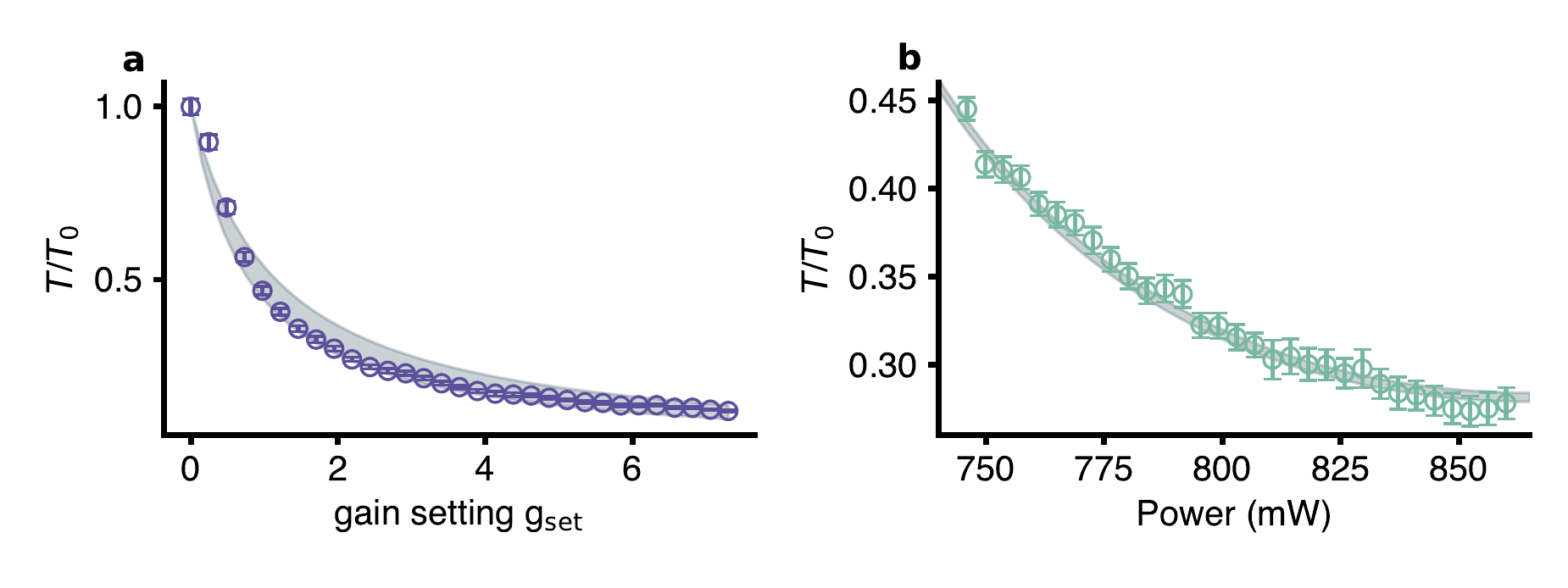}
\caption{ 
Temperature $T/T_0$ versus feedback gain for protocol $P_1$ (a) and laser trapping power for protocol $P_2$ (b) for the slowest realization ($\tau$= 2.26 ms), respectively. Data points are obtained from an ensemble average over 15000 repetitions of the protocol. The error bars correspond to statistical uncertainties from the finite ensemble size. We use Eq.~\eqref{eq:Teff_1} (Eq.~\eqref{eq:Teff_2}) to display the expectation in protocol $P_1$ ($P_2$) shown in a (b). The width of the corresponding grey band is determined by the uncertainty on the actual gain of the system $g$, respectively.}
\label{fig:calibration}
\end{figure*}

The effective temperature depends on the feedback gain $g$, which we control by our feedback circuit by setting a value $g_\text{set}$ (see Supplementary Note 3). The actual value of the gain $g$ does, however, not only depend on $g_\text{set}$, but also on slightly drifting parameters in the setup, like the position resolution of the detection scheme and the chamber pressure. We correspondingly determine the mean value of the gain $g$ and its relative uncertainty for a given set value $g_\text{set}$ independently, before and after the whole set of measurements for the respective protocols $P_1$ and $P_2$. The uncertainty in the actual gain $g$ is responsible for the uncertainty in setting the effective temperature. We compare this function with the data extracted from each protocol's quasistatic implementation. We use 15,000 repetitions to compute the temperature ratio $T/T_0$, where $T_0$ is the temperature with the feedback turned off, i.e., room temperature.

 For protocol $P_1$ the temperature ratio is given by:
 
  \begin{equation}
    T/T_0=1/\left(1+g\right).
    \label{eq:Teff_1}
\end{equation}
Here, the gain is increased linearly from $g=0$ to $g=7.3$. It has a relative uncertainty of $19\%$. Supplementary Figure \ref{fig:calibration}a shows the temperature ratio $T/T_0$ plotted versus the feedback gain setting $g_{set}$. The expectation shown as the grey band is based on Eq.~\eqref{eq:Teff_1} taking the uncertainty of the actual gain $g$ into account.

For protocol $P_2$, the laser power is tuned linearly from $750$~mW to $860$~mW, while the feedback gain is constant $g=2.6\pm0.1$. Supplementary Figure \ref{fig:calibration}b shows $T/T_0$ plotted versus the power of the trapping laser. The power dependence of the temperature ratio is given by the equation
\begin{equation}
\begin{aligned}
    T/T_0&=1/\left(1+g\sin\left(\sqrt{\frac{P}{P_0}}\frac{5\pi}{2}\right)\right)\\
    &=1/\left(1+g\sin\left(\frac{\omega}{\omega_0}\frac{5\pi}{2}\right)\right).
    \end{aligned}
    \label{eq:Teff_2}
\end{equation}
Again the band displays the range of theoretical expectation Eq.~\eqref{eq:Teff_2} when the uncertainty on the value of the gain $g$ is considered.



\begin{figure}[h]
\centering
\includegraphics[width=0.46\textwidth]{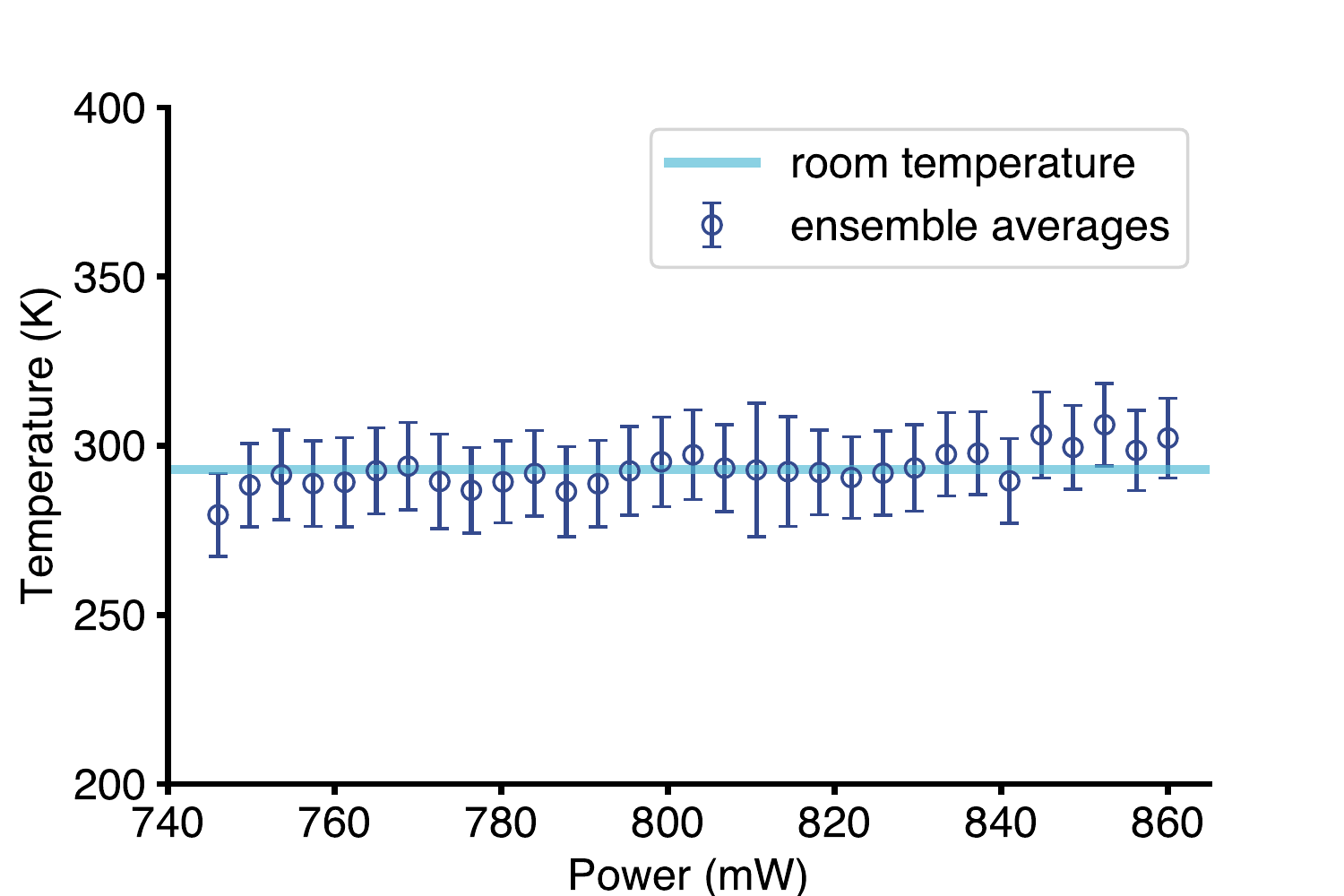}
\caption{Center-of-mass temperature of a $969$~nm diameter particle versus trapping laser power at a pressure of $3.5$ mbar. For 850 mW, the laser intensity is approximately 16 MW$/\text{center-of-mass}^2$. The temperature is determined from the variance of the particle trajectories with feedback laser off.}
\label{fig:powercheck}
\end{figure}
\begin{figure*}
\centering
\includegraphics[width=0.8\textwidth]{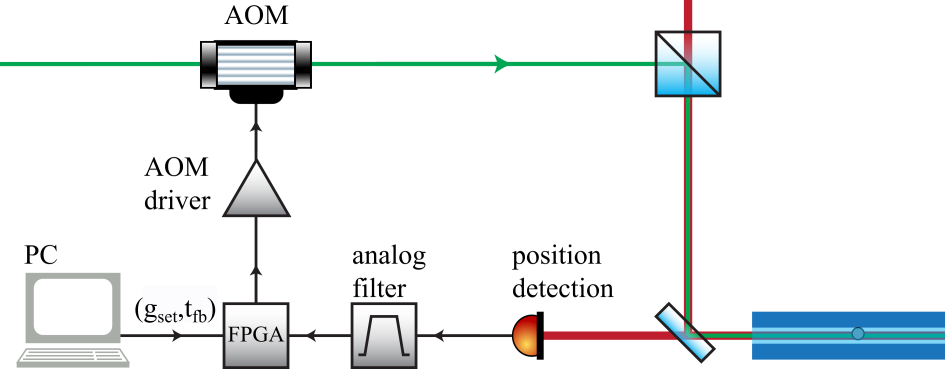}
\caption{Feedback loop. The position $x(t)$ of the particle is measured with a photodiode and bandpass filtered to suppress technical noise. The filtered signal is then processed by an FPGA: it introduces a delay by $t_\text{fb}$ and multiplies by a factor for feedback gain $g_\text{set}$. The resulting signal is sent to the modulation input of the AOM driver which creates the feedback force via radiation pressure on the particle.}
\label{fig:circuits}
\end{figure*}

\section{Influence of photon absorption on bath temperature} 
To investigate the fluctuation theorem (Eq.~(1) of the main text), we need to consider the effective bath temperature experienced by the center-of-mass motion of the levitated particle. In addition to the temperature of the surrounding gas and the effect of feedback cooling, it has been shown that the internal temperature of the microparticle couples to the center-of-mass motion and can also modify the bath temperature ~\cite{S_millen2014}. The internal temperature, in turn, may be increased by absorption of light from the optical tweezer. This absorption is typically higher in silica microparticles than in pure silica due to impurity of the material. 

While this has no influence on protocol $P_1$, where the trapping laser power is fixed, the power is varied during protocol $P_2$. Here we show an independent measurement to investigate the effect of heating by laser absorption on the bath temperature experienced by the microparticle's center-of-mass motion. To this end we vary the trapping power in the range [750-850] mW (consistently with protocol $P_2$) with feedback turned off. We determine the motional temperature of the particle via $\langle x^2\rangle=\frac{k_B T}{m\omega^2}$. The results are presented in Supplementary Figure~\ref{fig:powercheck}. The experimental data (blue circles, Supplementary Figure~\ref{fig:powercheck}) show no significant contribution of the laser power on the temperature of the center-of-mass motion. We conclude that additional heating due to light absorption is negligible in our experiment. This is not surprising, as we are operating at a pressure (approx.~$3.5$~mbar) where the internal temperature of the particle is still strongly cooled by collisions with the surrounding gas molecules.

\section{Description of the feedback control circuit} 

Supplementary Figure \ref{fig:circuits} provides a schematic description of the feedback control circuit. We observe the motion of the microparticle along the hollow-core photonic crystal fiber axis. The interference between the light scattered by the microparticle and the trapping laser results in a position dependent power monitored on a photo-detector (Thorlabs, PDB425C-AC). The resulting signal is filtered to suppress technical noise below (mainly of acoustic nature) and above (laser noise) the mechanical frequency. We choose an active, multiple feedback bandpass filter 
with a center frequency at $f_0=400$ kHz and a full width at half maximum bandwidth of $\Delta f=600$ kHz. 

The central element of our feedback circuitry is a field programmable gate array (FPGA) that allows real-time signal processing. We are using a PXIe-7965 (Virtex 5 based FPGA) in combination with a NI-5781 transceiver adapter module, both from National Instruments. Two control parameters are used to determine how the filtered signal is processed, specifically a time delay $t_\text{fb}$ and a factor to set the feedback $g_\text{set}$. To implement the two protocols, a higher level Labview program controls these parameters of the FPGA. For protocol $P_1$, the delay is constant $t_\text{fb}=\frac{5\pi}{2\omega_0}$ and the gain varies in time $g_{set}(t)$. For protocol $P_2$, the delay is set to $t_\text{fb}=\frac{5\pi}{2\omega_0}$ and the gain is fixed, which results in synchronous change of frequency and effective temperature (See supplement S1 for details on the gain setting). While the mechanical frequency $\omega$ is increased, the fixed delay time approaches the optimal value for cooling and the temperature of the center-of-mass motion of the particle is decreased. 

To apply the feedback force to the particle we use radiation pressure from a feedback laser. The output of the FPGA controls the power of an AOM driver,
 which then determines the power of our feedback laser (green in Supplementary Figure~\ref{fig:circuits}). 

The delay of the whole feedback circuitry has a minimum value. We determine this delay by sending a test signal to the FPGA instead of the detector signal. The test signal is pulsed, with each pulse composed of 4 oscillations of a square signal at a frequency of 400 kHz. The pulses are separated by 15~$\mu$s, exceeding by far the measured minimum delay. After passing the whole circuit, the signal from the feedback laser is recorded on the detector. We determine a minimum delay of 2.6~$\mu$s $=1.04\frac{2\pi}{\omega_0}$. To achieve the best possible feedback cooling at the frequency $\omega_0$, we introduce an additional delay of $0.49~\mu$s resulting in a total delay of $t_\text{fb}=\frac{5\pi}{2\omega_0}$. \\

\section{Thermal (or entropic) work}
\mo{Let us consider the  energy of the system $H(X,\kappa)$  with driving parameter $\kappa$ and  $X=(x,p)$.}
\mo{The infinitesimal change of the energy is given by \cite{seifert2012stochastic}
\begin{equation}
{d} H =  \frac{\partial H}{\partial X} {d} X + \frac{\partial H}{\partial \kappa} d\kappa = \delta Q + \delta W_\text{m},
\label{firstl}
\end{equation}
where $ \delta Q = \frac{\partial H}{\partial X} {d} X$ is the random heat along a single trajectory and $\delta W_\text{m} =  \frac{\partial H}{\partial \kappa} d\kappa = \frac{\partial H}{\partial \kappa} \dot \kappa dt$ is the stochastic mechanical work associated with a change of the external parameter $\kappa$. For a constant inverse temperature $\beta$, the dimensionless mechanical work, $W_\text{mech} = \beta W_\text{m}$, satisfies the fluctuation relation
 \cite{jarzynski1997nonequilibrium}
\begin{equation}
\langle \exp(- W_\text{mech}) \rangle = \exp(-\beta \Delta F).
\label{constTFT}
\end{equation}
For a time-dependent inverse temperature $\beta(t)$, a new contribution to the energy change has to be taken into account. The infinitesimal variation of the dimensionless energy $\beta H$ reads in this case,
\begin{equation}
{d} (\beta H) =  \beta \frac{\partial H}{\partial X} {d} X + \beta\frac{\partial H}{\partial \kappa} d\kappa + H \frac{d \beta}{d t} dt = \beta \delta Q + \delta W_\text{mech} + \delta W_\text{ther},
\label{firstl}
\end{equation}
with the dimensionless thermal (aka entropic) work $\delta W_\text{ther}= H\dot \beta dt$ associated with a change of the inverse temperature (or related entropy change) of the bath \cite{chelli2007generalization,chelli2007numerical,chatelain2007temperature,williams2008nonequilibrium,chelli2009nonequilibrium}. The fluctuation relation may accordingly be extended to \cite{chelli2007generalization,chelli2007numerical,chatelain2007temperature,williams2008nonequilibrium,chelli2009nonequilibrium}
\begin{equation}
\left\langle \exp(-[W_\text{mech}+W_\text{ther}] )\right\rangle = \exp[-\Delta(\beta {F})],
\label{eq:WSE}
\end{equation} 
with the variation of the dimensionless free energy $\Delta(\beta {F})$.}

\section{Clausius inequality} 
The fluctuation theorem (1) provides a nonequilibrium generalization of the Clausius inequality \cite{S_jarzynski1999}. Applying Jensen's inequality, we first obtain,
\begin{equation}
    \langle  {W}\rangle \geq \Delta (\beta  {F}).
    \label{work_ineq}
\end{equation}
Using the first law of thermodynamics along a nonequilibrium  protocol, we further have  $\Delta (\beta \langle H\rangle)= \langle {W}\rangle+ \langle \int \beta \delta {Q}\rangle_\text{neq}$, where $\Delta (\beta \langle H\rangle)=\beta(\tau) \langle H(\tau)\rangle -\beta(0) \langle H(0)\rangle$ and $Q$ in the heat exchanged with the bath at inverse temperature $\beta$. On the other hand, from the definition of the equilibrium free energy difference, we have $\Delta (\beta  {F})= \beta(\tau) \langle H(\tau)\rangle_\text{eq} -\beta(0) \langle H(0)\rangle_\text{eq} -\Delta S$, where $S$ is the (dimensionless) equilibrium entropy of the system. We obtain accordingly,
\begin{equation}
 \left \langle \int \beta \delta {Q}\right \rangle_\text{neq}\leq \Delta (\beta \langle H\rangle)-\Delta (\beta \langle H\rangle_\text{eq})+ \Delta S.
\label{heat_ineq}
\end{equation}
For the considered  harmonic oscillator,  equipartition implies that the (dimensionless) equilibrium energy difference  vanishes, $\Delta (\beta \langle H\rangle_\text{eq})=0$. The  Clausius inequality for a nonequilibrium protocol followed by an isothermal thermalization thus reads,
\begin{equation}
\left \langle \int \beta \delta {Q}\right \rangle = \left \langle \int \beta \delta {Q}\right \rangle_\text{neq} - \Delta (\beta \langle H\rangle)\leq \Delta S.
\label{heat_ineq}
\end{equation}

\section{Distribution of mechanical work}

In Supplementary Figure~\ref{fig:pdf_Wm}, we show the probability distribution function for the mechanical work $W_\text{mech}$ corresponding to 15,000 repetitions of protocol $P_2$ (see Fig. 2 of main text for more details). For fast driving (left) the distribution is clearly asymmetric while for slow, quasistatic driving, it  is close to a Gaussian distribution, as expected. 

\begin{figure}[h]
\centering
\includegraphics[width=0.48\textwidth]{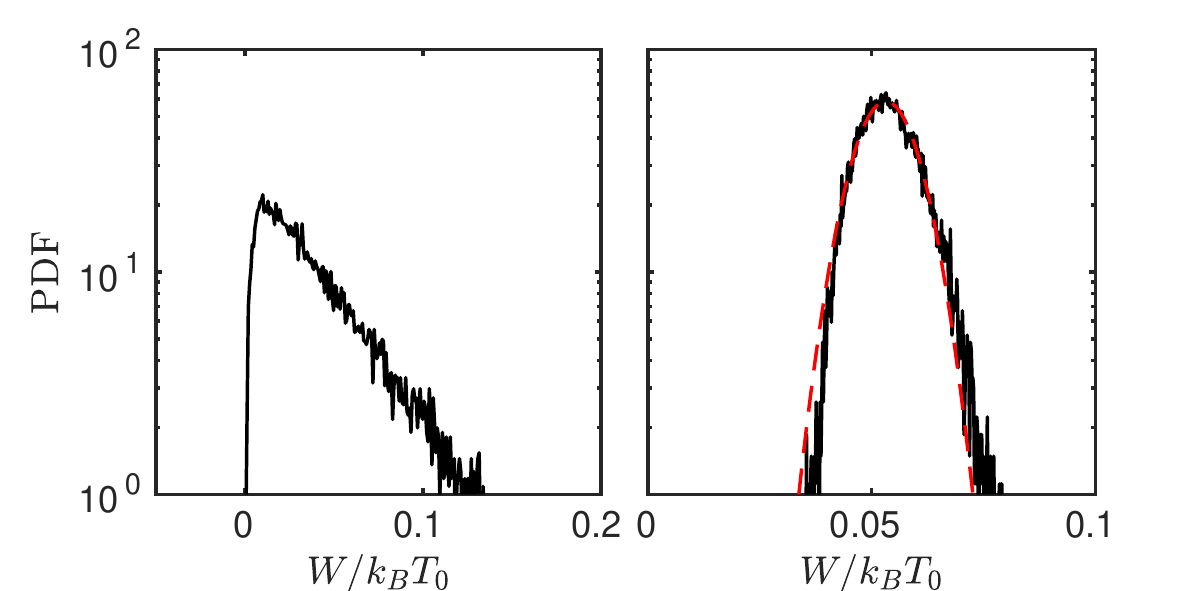}
\caption{Distribution function of mechanical work $W_\text{mech}$ for 15,000 repetitions of protocol $P_2$. Left: fast driving ($\tau=22.6~\mu s$). Right: slow quasistatic driving ($\tau=2.26$ ms). The dashed red line is the best Gaussian fit.}
\label{fig:pdf_Wm}
\end{figure}


%

\end{document}